# Title: A Radio Counterpart to a Neutron Star Merger

**Authors:** G. Hallinan[1*‡], A. Corsi[2‡], K. P. Mooley[3], K. Hotokezaka[4,5], E. Nakar[6], M.M. Kasliwal[1], D.L. Kaplan[7], D.A. Frail[8], S.T. Myers[8], T. Murphy[9,10], K. De[1], D. Dobie[9,10,11], J.R. Allison[9,12] K.W. Bannister[11], V. Bhalerao[13], P. Chandra[14]†, T.E. Clarke[15], S. Giacintucci[15], A.Y.Q. Ho[1], A. Horesh[16], N.E. Kassim[15], S. R. Kulkarni[1], E. Lenc[9,10], F. J. Lockman[17], C. Lynch[4,10], D. Nichols[18], S. Nissanke[18], N. Palliyaguru[2], W.M. Peters[9], T. Piran[12], J. Rana[19], E. M. Sadler[9,10], L.P. Singer[20]

**Affiliations:**
[1]Division of Physics, Mathematics and Astronomy, California Institute of Technology, 1200 East California Boulevard, Pasadena, CA 91125, USA
[2]Department of Physics and Astronomy, Texas Tech University, Box 41051, Lubbock, TX 79409-1051, USA
[3]Astrophysics, Department of Physics, University of Oxford, Keble Road, Oxford OX1 3RH,UK
[4]Center for Computational Astrophysics, Flatiron Institute, 162 5th Ave, New York, NY 10010, USA
[5]Department of Astrophysical Sciences, Princeton University, Peyton Hall, Princeton, NJ 08544 USA
[6]The Raymond and Beverly Sackler School of Physics and Astronomy, Tel Aviv University, Tel Aviv 69978, Israel
[7]Department of Physics, University of Wisconsin, Milwaukee, WI 53201, USA
[8]National Radio Astronomy Observatory, Socorro, New Mexico, 87801, USA
[9]Sydney Institute for Astronomy, School of Physics, The University of Sydney, NSW 2006, Australia
[10]Australian Research Council Centre of Excellence for All-sky Astrophysics (CAASTRO)
[11]Australia Telescope National Facility, Commonwealth Scientific and Industrial Research Organisation, Astronomy and Space Science, PO Box 76, Epping, NSW 1710, Australia
[12]Australian Research Council Centre of Excellence for All-sky Astrophysics in 3 Dimensions (ASTRO 3D)
[13]Department of Physics, Indian Institute of Technology Bombay, Mumbai 400076, India
[14]National Centre for Radio Astrophysics, Tata Institute of Fundamental Research, Pune University Campus, Ganeshkhind Pune 411007, India
[15]Remote Sensing Division, Naval Research Laboratory, Code 7213, 4555 Overlook Ave. S W, Washington, DC 20375
[16]Racah Institute of Physics, The Hebrew University of Jerusalem, Jerusalem, 91904, Israel
[17]Green Bank Observatory, P.O. Box 2, Green Bank, WV 24944
[18]Institute of Mathematics, Astrophysics and Particle Physics, Radboud University, Heyendaalseweg 135, 6525 AJ Nijmegen, The Netherlands
[19]Inter University Centre for Astronomy and Astrophysics (IUCAA), S. P. Pune University Campus, Pune, Maharashtra 411007, India
[20]Astroparticle Physics Laboratory, NASA Goddard Space Flight Center, Mail Code 661, Greenbelt, MD 20771, USA

† Current address: Department of Astronomy, Stockholm University, Alba Nova, SE-106 91 Stockholm

* Correspondence to: gh@astro.caltech.edu

‡ These authors contributed equally to this work.

**Abstract**: Gravitational waves have been detected from a binary neutron star merger event, GW170817. The detection of electromagnetic radiation from the same source has shown that the merger occurred in the outskirts of the galaxy NGC 4993, at a distance of 40 megaparsecs from Earth. We report the detection of a counterpart radio source that appears 16 days after the event, allowing us to diagnose the energetics and environment of the merger. The observed radio emission can be explained by either a collimated ultra-relativistic jet viewed off-axis, or a cocoon of mildly relativistic ejecta. Within 100 days of the merger, the radio light curves will distinguish between these models and very long baseline interferometry will have the capability to directly measure the angular velocity and geometry of the debris.

**One Sentence Summary:** The radio counterpart of the binary neutron star merger GW170817 probes the energetics and environment of the explosion.

## Main text

On 2017 August 17, the Advanced Laser Interferometer Gravitational Wave Observatory (Advanced LIGO) detected a gravitational wave signal, GW170817, which was rapidly identified to be associated with the inspiral and coalescence of two neutron stars (*1*). A burst of gamma-rays, GRB170817A, was detected approximately two seconds after the gravitational wave detection by the Gamma-ray Burst Monitor (GBM) of the Fermi Gamma-ray Space Telescope (*2*–*4*). With the addition of data from the Advanced Virgo interferometer, the source of gravitational waves was localized to an area of 28 deg$^2$ (90% confidence region) and a distance of 40 ± 8 megaparsecs (Mpc) (*1*). There were 49 cataloged galaxies within this volume, allowing astronomers to rapidly search for electromagnetic counterparts (5). An optical counterpart, designated SSS17a, was detected within ~11 hours of the event by astronomers using the Swope telescope, localizing the merger to the S0-type galaxy NGC 4993 at a distance of 40 Mpc *(6,7)*. It was independently confirmed soon after (*8,9*). Following the optical detections, targeted observing campaigns were initiated across the electromagnetic spectrum (*10*). Subsequent optical and infrared spectroscopic observations firmly established this optical counterpart to be associated with the neutron star merger GW170817 (*5*).

We report a coordinated effort to use the Karl G. Jansky Very Large Array (VLA), the VLA Low Band Ionosphere and Transient Experiment (VLITE), the Australia Telescope Compact Array (ATCA) and the Giant Metrewave Radio Telescope (GMRT) to constrain the early time radio properties of the neutron star merger. Companion papers report the ultraviolet and X-ray properties (*11*) and interpret the panchromatic behavior of the transient (*5*). The multi-wavelength counterpart to GW170817 is hereafter referred to as EM170817.

## The Search for a Radio Counterpart to GW170817

We began radio observations of NGC 4993 on August 17 2017 at 01:46 UTC, within ~13 hours of the detection of the gravitational event. These initial observations were part of a survey with the ATCA, targeting galaxies in the gravitational wave localization region as identified by the Census of the Local Universe (CLU) catalog (*5*). A similar survey of these CLU cataloged galaxies also commenced with the VLA. After confirmation of a compelling optical counterpart to the merger, observations focused on the location of EM170817, with coordination between the VLA, ATCA, GMRT and VLITE enabling monitoring on a close to daily basis at frequencies spanning 0.3 – 10

GHz (*12*). Only upper limits on the radio flux of EM170817 were possible until a counterpart appeared in VLA data from observations on September 2 and September 3 at a frequency of 3 GHz, and in independent observations on September 3 at a frequency of 6 GHz (Figure 1) (*13*, *14*). The ATCA also detected the source on September 5 in the 5.5 – 9 GHz band (*15*). The entire radio data set is tabulated in Table S1. In observations at 3 GHz with the VLA, the source shows evidence of an increase in flux density over a timescale of 2 weeks, varying from 15.1 ± 3.9 μJy on 2017 September 03 to 34 ± 3.6 μJy (1 Jy = $10^{-23}$ erg $s^{-1}$ $cm^{-2}$ $Hz^{-1}$) on 2017 September 17 (Figures 2 and 3).

Figure 1 shows a comparison between a near-infrared image of EM170817 and deep radio images of the same field from the VLA at a frequency of 6 GHz. The position of the radio source is Right Ascension (RA) = 13h09m48.061s ± 0.005s and Declination (Dec) = -23d22m53.35s ± 0.14s (J2000 equinox), using data at 3 GHz from September 8 and 10. The optical position derived from Hubble Space Telescope observations is RA = 13h09m48.071s ± 0.004s, Dec = -23d22m53.37s ± 0.05s (*16*). Within the uncertainties, the two positions are mutually consistent. We can further calculate the chance alignment of the optical counterpart to a background radio source. In these same radio data (Sept 8 and 10 combined), the radio source has reached a flux density of 25 ± 2.2 μJy. There are 2,700 sources per square degree with flux density greater or equal to this value at 3 GHz (*17*), resulting in a likelihood of chance alignment of $2 \times 10^{-5}$ for the positional errors shown above. The likelihood of chance alignment becomes even smaller when considering that the source has been observed to double in flux density over 2 weeks and <4% of sources at 3 GHz vary by >30% (*18*). We therefore confirm the transient to be the radio counterpart to EM170817.

Models of binary neutron star coalescence predict the emergence of an associated radio flare, due to the tidal ejection of 0.01-0.05 solar masses ($M_\odot$) of energetic material at sub-relativistic velocities (a few tenths of the speed of light) (*19 – 21*). This ejecta material forms a blast wave as it plows through the ambient interstellar medium (ISM) surrounding the merger, producing synchrotron radiation peaking at radio frequencies, lasting months to years after the merger. The observed timescale and luminosity of the radio source is sensitive to the mass and velocity of the blast wave ejecta and the density of the ISM. Therefore, the radio emission is diagnostic of the energetics of the blast wave ejecta as well as the environment of the merger.

Binary neutron star mergers have long been proposed as a likely progenitor for short (< 2 s) hard gamma-ray bursts (sGRBs) (*22*) and the generation of an ultra-relativistic jet is required to account for the properties of both the GRBs and their afterglows (*23*). As in the case of sub-relativistic ejecta, the jet interacting with the circum-merger medium will produce radio emission. However, in this case, the resulting radio light curve depends critically on the angle between the observer line of sight and the jet (*24*).

We compare our early time radio observations with numerical models (*12*) for the expected radio light curves due to synchrotron emission from sub-relativistic ejecta and an ultra-relativistic jet, as well as the interaction between these components. Where possible, we focus on a portion of the parameter space that is consistent with observations at optical, infrared, UV and X-ray [see (*5*), their figure 5, for a schematic illustration of the model].

## The Sub-Relativistic Ejecta

The radio flare produced by the sub-relativistic ejecta that generates the optical and infrared emission is expected to peak on a timescale of months to years (*19–21*). With the mass and velocity range inferred from the optical and infrared, its energy must be high (>$10^{51}$erg) (*5*); however, with an expected velocity of ~0.2c the bulk of this ejecta cannot be the source of the radio signal that has been observed to rise within weeks of the merger. Instead, this component is expected to dominate the radio emission at late time (years). Nevertheless, this ejecta is expected to have a distribution of velocities with a low-mass fast tail that can extend up to mildly relativistic velocities, which may be the source of the observed emission (Figure 3). When considering the sub-relativistic ejecta, we take into account the non-detection of neutral hydrogen from the host galaxy (5σ mass limit of <$1 \times 10^{8}$ $M_\odot$) in our recent observations with the Green Bank Telescope (GBT, *12*), which suggests an ISM density, $n$<0.04 $cm^{-3}$ (supplementary online text). Assuming this density as a constraining limit, the radio then requires such a mildly relativistic outflow to have a velocity v>0.7c (Γ>1.5) and to carry at least $10^{49}$ erg in isotropic equivalent energy (Figure 3). This is unlikely, but cannot be ruled out.

## A Classical Short Hard Gamma Ray Burst

GW170817 provides an unambiguous detection of a binary neutron star merger and therefore offers the opportunity to directly investigate the presence of an ultra-relativistic jet. The isotropic equivalent luminosity of the burst of gamma-rays detected by Fermi GBM is $4 \times 10^{46}$ erg (*2–4*). This is orders of magnitude lower than the peak luminosity of the classical sGRB population ($10^{49}$ – $10^{52}$ erg; median = $2 \times 10^{51}$ erg) (*23*). Moreover, this gamma-ray emission is not a very low luminosity analog of the classical sGRB population (*5*).

Therefore, the observed gamma-rays cannot securely confirm the long-standing hypothesis that NS mergers are the progenitors of cosmological sGRBs. Within the framework of the classical short hard GRB model, there are two possibilities: i) the jet axis was slightly offset from our line-of-sight, but was close enough for the observed gamma-rays to be a component of the regular sGRB prompt emission (hereafter called the slightly off-axis model); or ii) our line-of-sight was at a large angle from the jet axis and the observed gamma-rays were generated by a different mechanism (hereafter called the widely off-axis model). Below we explore the current radio constraints and predict the future evolution under each of these scenarios.

In the slightly off-axis model, if the gamma-rays are produced by a slightly off-axis jet, then the edge of the jet cannot be more than about 0.1 rad from our line of sight (*5*). The jet drives a relativistic blast wave into the ISM, which subsequently decelerates quickly. The Lorentz factor Γ drops to ~ 10 within about a day, after which the beam of its emission expands to include our line of sight, thereby producing bright radio emission in our direction. The blast wave Lorentz factor at a given time depends very weakly on the jet isotropic equivalent energy $E_{iso}$ and the external number density $n$; therefore this prediction holds for a wide range of jet and ISM parameters (*25*).

Figure 2 shows several predicted light curves (*12*) for an ultra-relativistic jet with $E_{iso}$=$10^{50}$ erg that misses our line of sight by 0.1 rad. A density of $10^{-3}$ $cm^{-3}$, which is on the low end of the distribution of densities inferred from sGRB afterglows (*26*), produces a signal that is brighter by more than an order of magnitude than the observed radio emission from EM170817. A very low density of $6 \times 10^{-7}$ $cm^{-3}$ is required to reproduce the observed light curve, which is more consistent with the intergalactic medium than the environs of a S0-type galaxy (*27*). Therefore, our radio observations strongly disfavor a model in which the observed gamma-rays were produced by a slightly off-axis luminous ultra-relativistic jet.

Under the widely off-axis model, the radio data are consistent with the model for reasonable ranges of jet energy and ISM density (Figure 3). The radio emission from an off-axis jet undergoes a rapid rise before a broad peak, followed by a slow decline. The observed radio light curve of EM170817 is inconsistent with the sharp rise phase of an off-axis jet. Instead, the light curve implies that the observations were made around the onset of the broad peak in emission. Thus, a prediction of this model is that the radio light curve will remain at a similar brightness for the next several weeks and then start to fade. However, this model does not account for the early-time gamma-ray emission.

**A Mildly Relativistic Cocoon**

The high luminosity of the optical and IR counterpart to EM170817 requires a high mass of ejecta at sub-relativistic velocities at the time of the merger (*5*). The observed delay between the gravitational wave signal of the neutron star merger and the Fermi-detected gamma rays indicates an additional sustained source of energy post-merger. This source of energy is likely manifested as a jet. As it expands, the jet transfers a large fraction of its energy into the surrounding ejecta, forming a hot cocoon that expands over a wide angle whilst traveling at mildly relativistic velocities (*28*). The formation and possible eventual break-out of such a cocoon can account for many of the properties of the observed electromagnetic signatures seen after the event, from infrared to gamma-rays *(5)*. As this cocoon propagates into the ISM, it will also produce a radio signal.

The energy coupled to the mildly relativistic ejecta depends mostly on the fate of the jet, as well as its total energy. If the jet is narrow (opening angle ~10°), it can drill through the ejecta and break out with an isotropic equivalent energy of $\sim 10^{50} - 10^{51}$ erg, leaving a fraction of its total energy ($10^{48} - 10^{49}$ erg) in the mildly relativistic cocoon. Alternatively, if the jet is wide it requires about $10^{51}$ erg in order to propagate a substantial distance within the ejecta before it is fully choked, depositing all of its energy in the cocoon. A non-negligible fraction of this energy, $>10^{50}$ erg, is then coupled to mildly relativistic ejecta with $\Gamma = 2 - 3$ (and possibly even higher). Thus an energetic, mildly relativistic outflow indicates a choked jet, while a low-energy, mildly relativistic outflow indicates a narrow jet that breaks out of the ejecta.

Figure 3 shows the predicted radio emission for both scenarios (*12*), assuming $\Gamma = 2$, showing that they are consistent with the observed radio emission if the ISM density is about $\sim 3 \times 10^{-3}$ $cm^{-3}$ ($\Gamma = 3$ requires a density of $\sim 3 \times 10^{-4}$ $cm^{-3}$). Both curves are similar during the rising phase, as the radio emission is generated by the forward shock propagating into the ISM with luminosity that depends only on velocity (for a given density). This velocity is constant during the rising phase. During this phase, the radio flux at a given band rises as time $t^3$ over a wide range of parameters, regardless of the shock velocity (whether Newtonian or relativistic), cocoon energy or ISM density. The emission peaks when the entire energy of the cocoon is deposited in the ISM and the shock starts to decelerate. Thus, while an energetic cocoon may become much brighter several months after a merger, a low-energy cocoon persists for much less time. The current data marginally favor the low-energy cocoon model (jet break-out) over the energetic cocoon (choked jet). We predict that the radio light curves will definitively discriminate between an energetic cocoon, a low-energy cocoon and an off-axis jet model within 100 days of the merger.

**Consistency with the X-ray observations**

All the models that we have considered predict an optically thin spectrum, consistent with our radio observations (Figure SI4), with a single power-law between the radio and the X-rays, $F_\nu \propto \nu^{-\beta}$, where $F_\nu$ is flux density, $\nu$ is frequency and $\beta$ is the spectral power law index. $\beta$ depends on the electron distribution power-law index, *p*, as $\beta=(p-1)/2$. Thus, all our models predict the same X-ray flux around the time that radio emission was detected. This flux is broadly consistent with the X-ray measurement at 15 days (*29,30*) if $\beta=0.5$ ($p=2$). This spectral index and value of p are lower than typically observed in GRB afterglows, although there are afterglows where such values are measured (*31*). If $p>2$ then the models of the radio emission predict an X-ray flux that is lower than the observed one. In that case, the observed X-rays must not be produced by the blast wave that propagates into the ISM, but by some other source.

**Imaging the Fireball of GW170817**

Our models predict differing sizes for the expanding radio emission region (Figure 4). Radio observations can directly and indirectly constrain the size of the expanding fireball, as has been demonstrated in the case of long GRBs (*32*, *33*).

Radio sources of compact size can be observed to vary, sometimes by a large degree on short timescales, due to interstellar scintillation on propagation through the ISM of our own Galaxy. This variability, analogous to the twinkling of compact objects observed at optical wavelengths, can be used to indirectly measure the size of compact radio sources (*34*). Using a simple model of the ISM in our Galaxy, inferred from observations of pulsars (*35*), we predict that the radio counterpart to EM170817 will be subject to refractive scintillation in the strong scattering regime. We calculate the expected modulation index and characteristic timescale for scintillation of the various possible components of ejecta (Figure S5). We find the degree of modulation unlikely to be useful in constraining the source size, given the low signal-to-noise ratio of the radio detections, except in the case of sub-relativistic ejecta. Conversely, this suggests that the light curves presented in Figures 2 and 3 are reliable measures of the intrinsic variability of EM170817, not misidentified scintillation.

A more direct method to constrain the size of the afterglow and directly measure the outflow front velocity is Very Long Baseline Interferometry (VLBI), which has been successfully applied to the case of long GRBs (*33*). We predict that EM170817 will become detectable and resolved on VLBI baselines within 100 days of the merger, providing an independent constraint on the nature of the ejecta (Figure 4).

**Acknowledgements:**

GH, AC and KPM would like to acknowledge the support and dedication of the staff of the National Radio Astronomy Observatory and particularly thank the VLA Director, Mark McKinnon, as well as Amy Mioduszewski and the VLA Schedulers, for making the VLA campaign possible. The National Radio Astronomy Observatory is a facility of the National Science Foundation operated under cooperative agreement by Associated Universities, Inc. We thank the staff of the GMRT that made these observations possible. The GMRT is run by the National Centre for Radio Astrophysics of the Tata Institute of Fundamental Research. The Australia Telescope Compact Array is part of the Australia Telescope National Facility which is funded by the Australian Government for operation as a National Facility managed by CSIRO. We thank the Green Bank Observatory for their rapid response to our Director's Discretionary Time GBT proposal. The Green Bank Observatory is a facility of the National Science Foundation operated under a cooperative agreement by Associated Universities, Inc. GH acknowledges the support of NSF award AST-1654815. A.C. and N.T.P acknowledge support from the NSF award #1455090 "CAREER: Radio and gravitational-wave emission from the largest explosions since the Big Bang". K.P.M's research is supported by the Oxford Centre for Astrophysical Surveys which is funded through the Hintze Family Charitable Foundation. This work was supported by the GROWTH (Global Relay of Observatories Watching Transients Happen) project funded by the National Science Foundation under PIRE Grant No 1545949. GROWTH is a collaborative project among California Institute of Technology (USA), University of Maryland College Park (USA), University of Wisconsin Milwaukee (USA), Texas Tech University (USA), San Diego State University (USA), Los Alamos National Laboratory (USA), Tokyo Institute of Technology (Japan), National Central University (Taiwan), Indian Institute of Astrophysics (India), Inter-University Center for Astronomy and Astrophysics (India), Weizmann Institute of Science (Israel), The Oskar Klein Centre at Stockholm University (Sweden), Humboldt University (Germany), Liverpool John Moores University (UK). EN acknowledges the support of an ERC starting grant (GRB/SN) and an ISF grant (1277/13). This work is part of the research program Innovational Research Incentives Scheme (Vernieuwingsimpuls), which is financed by the Netherlands

Organization for Scientific Research through the NWO VIDI Grant No. 639.042.612-Nissanke and NWO TOP Grant No. 62002444--Nissanke. PC acknowledges support from the Department of Science and Technology via SwarnaJayanti Fellowship awards (DST/SJF/PSA-01/2014-15). AH acknowledges support by the I-Core Program of the Planning and Budgeting Committee and the Israel Science Foundation. TM acknowledges the support of the Australian Research Council through grant FT150100099. Parts of this research were conducted by the Australian Research Council Centre of Excellence for All-sky Astrophysics in 3D (ASTRO 3D) through project number CE170100013. A.Y.Q.H. was supported by a National Science Foundation Graduate Research Fellowship under Grant No. DGE-1144469. Parts of this research were conducted by the Australian Research Council Centre of Excellence for All-sky Astrophysics (CAASTRO), through project number CE110001020. DLK is supported by NSF grant AST-1412421. TP acknowledges the support of Advanced ERC grant TReX. Basic research in radio astronomy at the Naval Research Laboratory (NRL) is funded by 6.1 Base funding. Construction and installation of VLITE was supported by NRL Sustainment Restoration and Maintenance funding. VB acknowledges the support of the Science and Engineering Research Board, Department of Science and Technology, India, for the GROWTH-India project. The VLA data reported in this paper are made available through the NRAO data access portal at http://doi.org/XXXXXXX.

**Supplementary Materials:**

Materials and Methods

Supplementary Text

Figures S1-S5

Table S1

References (*36*-*65*)

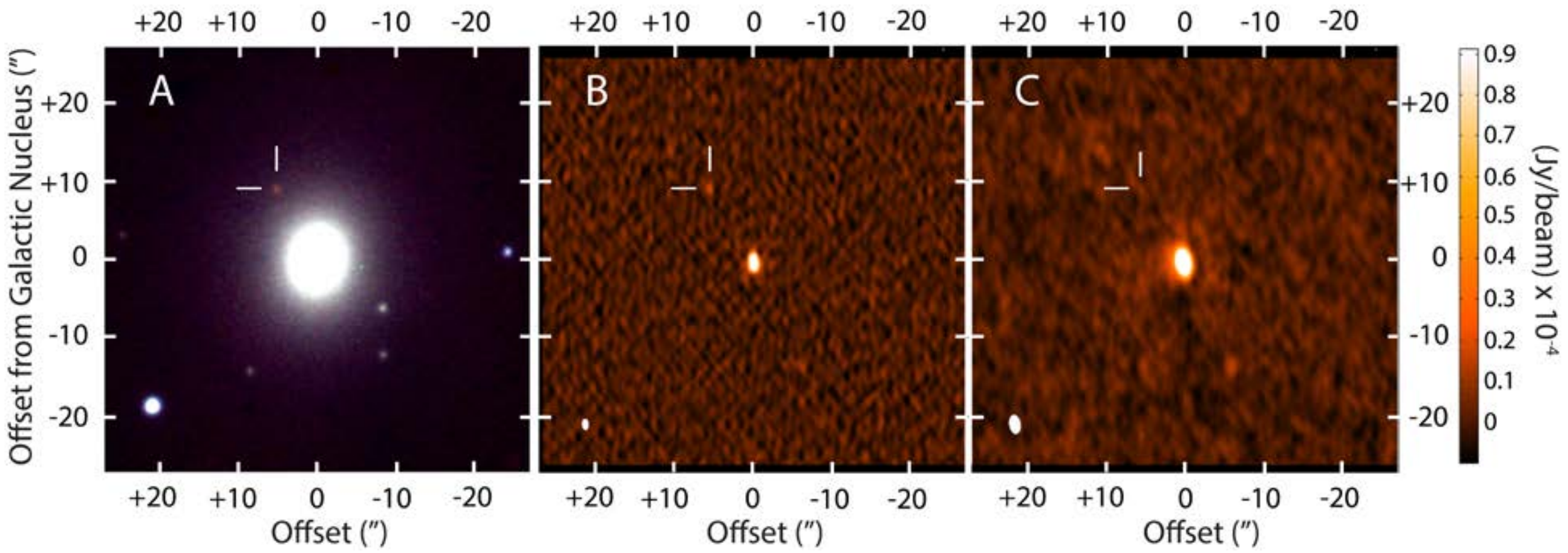

**Fig. 1**. **Comparison of the near-infrared and radio counterparts to EM170817**. (A) Near-infrared image of NGC 4993 with EM170817 highlighted, assembled from *J*, *H*, and *Ks* photometric bands taken with the FLAMINGOS-2 instrument on Gemini-South on 2017 August 27 (*3*). (B) Radio image of the same field created using VLA observations (6 GHz) on 2017 September 9, with the radio counterpart to EM170817 highlighted. Its flux density is 23 ± 3.4 μJy. (C) A combined image from four VLA observations at 6 GHz spanning 2017 August 22.6 – September 1. The flux density at the position of EM170817 is 7.8 ± 2.6 μJy, consistent with a marginal or non-detection. Radio emission seen in (B) and (C) from the core of the galaxy is due to an active galactic nucleus (AGN).

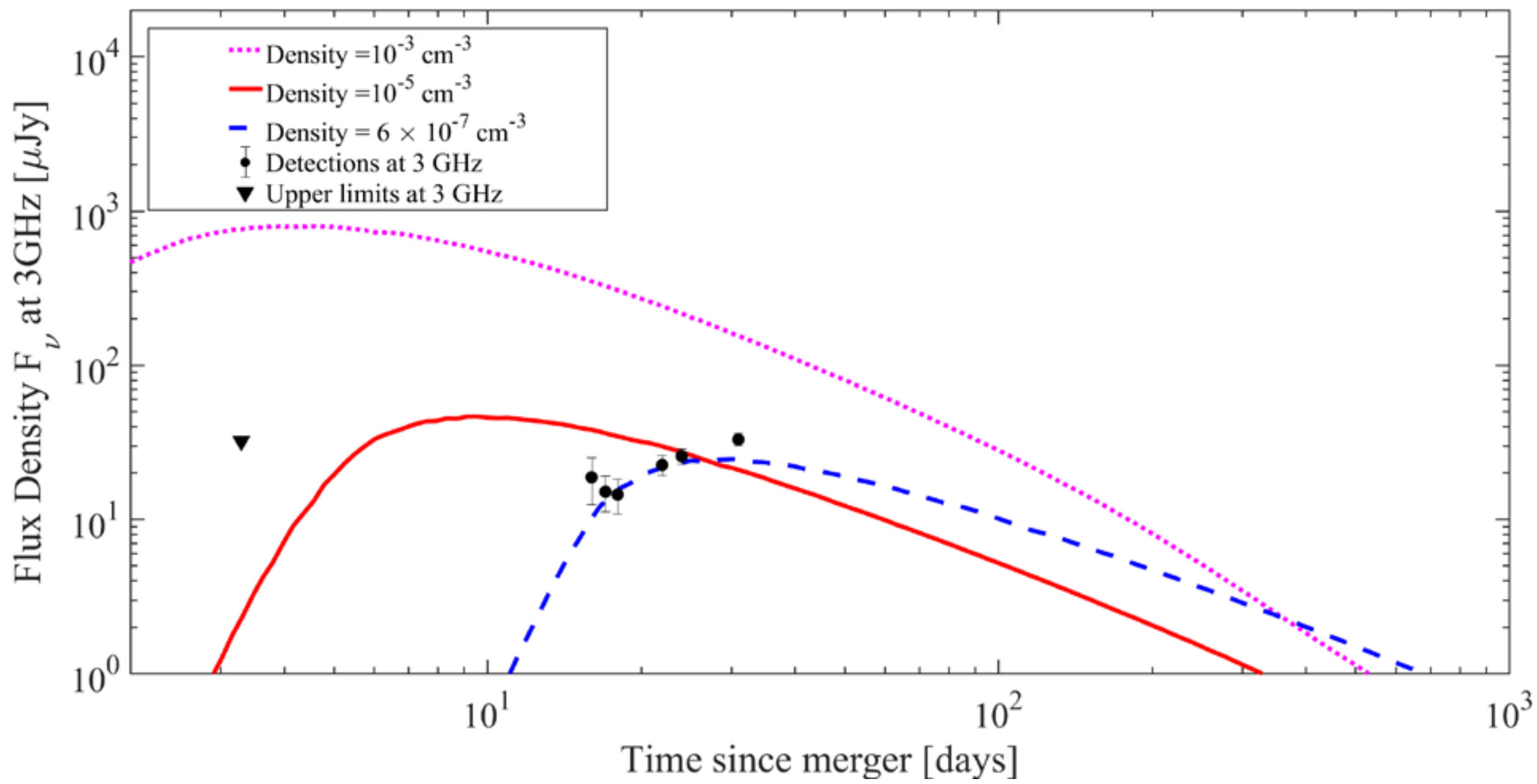

**Fig. 2. Radio observations rule out a slightly off-axis jet**. Invoking a slightly off-axis jet to explain the Fermi-detected gamma-rays from EM170817 (*3*) would require an associated radio afterglow appearing within a few days. The radio counterpart to EM170817 is inconsistent with this model. Light curves are shown for three examples, each with a jet isotropic equivalent energy, $E_{iso}=10^{50}$ erg, a jet half-opening angle of ~25° and an angle between our line of sight and the jet of ~30°. The only parameter that varies between them is density. The light curves associated with a density of $10^{-3}$ $cm^{-3}$ (pink dotted curve) and $10^{-5}$ $cm^{-3}$ (red solid curve) are completely inconsistent with the data. The data can be fitted by a model of a jet ($E_{iso}$ = 4 x $10^{50}$ erg) interacting with an ISM density of $6 \times 10^{-7}$ $cm^{-3}$ (blue dashed curve) inconsistent with the ISM density of a galaxy.

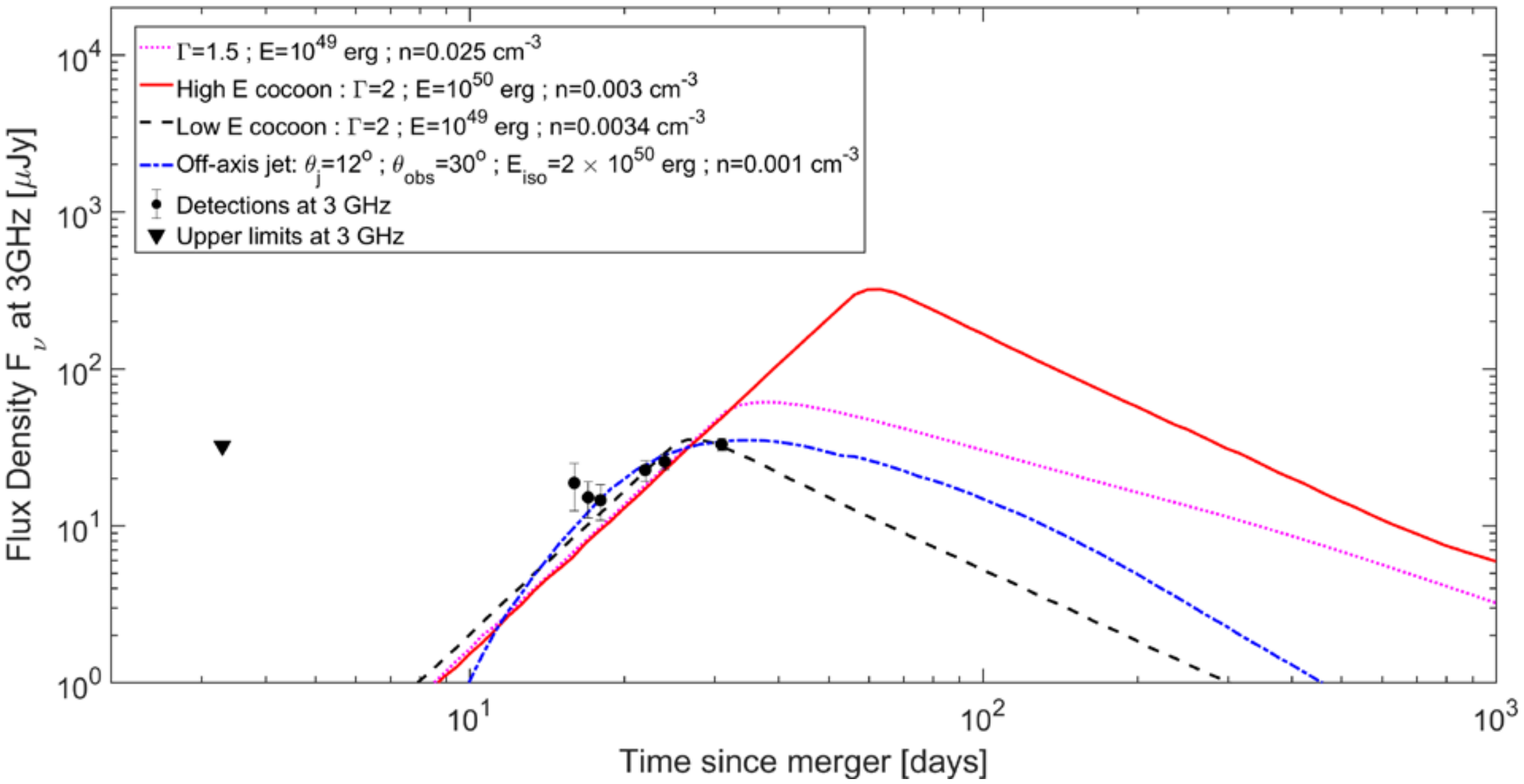

**Fig. 3. Radio light curve is consistent with either an off-axis jet or a cocoon.** Light curves are shown for both proposed cocoon models, i.e. a high energy cocoon due to a choked jet (red solid curve) and a low energy cocoon with jet break-out (black dashed curve). The light curve favors the low energy cocoon with jet break-out. For the off-axis jet, we use a jet isotropic equivalent energy, $E_{iso} = 1.5 \times 10^{50}$ erg, a jet half-opening angle, $\theta_j \sim 12°$, and an angle between our line of sight and the jet, $\theta_{obs} \sim 30°$, together with an ISM density of $10^{-3}$ $cm^{-3}$ (blue dash-dot curve). However, the light curve is consistent with a range of parameter space in jet energy and ISM density. For each set of jet and observer angle there is a single solution (namely *n* and *E*) that fit the data, which cannot be distinguished from the light curve. The pink curve is a model that represents the expected radio emission due to the high velocity tail of the sub-relativistic ejecta, which is yet to be ruled out. All the models indicate an ISM density of $\sim 10^{-3} - 10^{-2}$ $cm^{-3}$, which in turn is consistent with constraints from H I upper limits (supplementary online text). We predict that the radio light curves will distinguish between the models shown within the first 100 days of the merger.

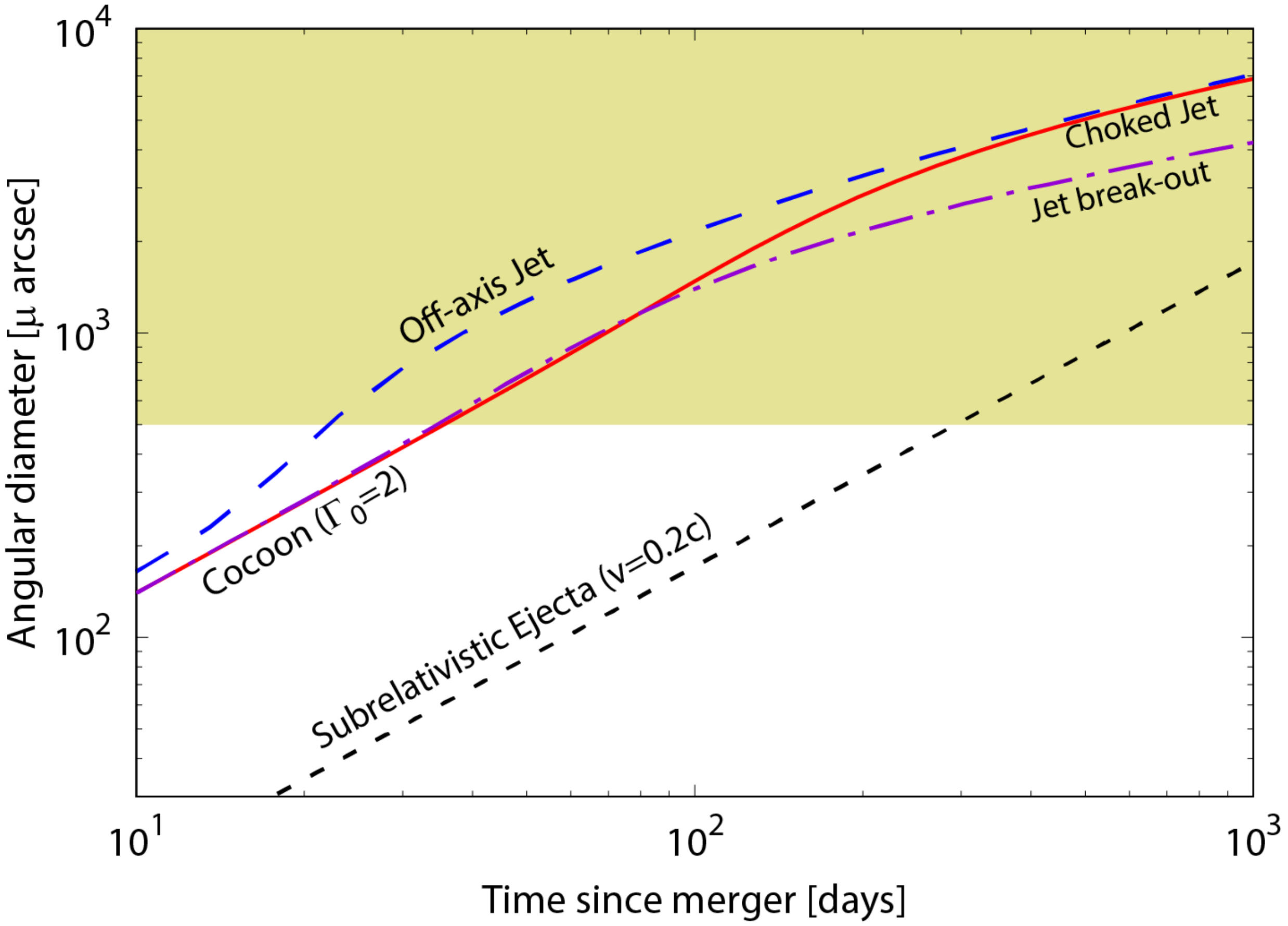

**Fig. 4. Predicted time evolution of the radio source size for different models.** Different possible models for the radio emission from EM170817 expand at different velocities. The yellow area shows the approximate parameter space accessible to VLBI. Model parameters for the off-axis jet and cocoon models are the same as used in Figure 3. The sub-relativistic ejecta is assumed to have a bulk velocity of 0.2*c*, as discussed in the main text. We note that this represents a conservative lower limit to the source size as the radio emission is dominated by the fastest component of the ejecta which can exceed 0.4*c*.

# Supplementary Materials for

## A Radio Counterpart to a Neutron Star Merger

G. Hallinan[1*‡], A. Corsi[2‡], K. P. Mooley[3], K. Hotokezaka[13,20], E. Nakar[15], M.M. Kasliwal[1], D.L. Kaplan[14], D.A. Frail[11], S.T. Myers[11], T. Murphy[4,10], K. De[1], D. Dobie[4,6,10], J.R. Allison[4,5] K.W. Bannister[6], V. Bhalerao[7], P. Chandra[8]†, T.E. Clarke[9], S. Giacintucci[9], A.Y.Q. Ho[1], A. Horesh[12], N.E. Kassim[9], S. R. Kulkarni[1], E. Lenc[4,10], F. J. Lockman[19], C. Lynch[4,10], D. Nichols[16], S. Nissanke[16], N. Palliyaguru[2], W.M. Peters[9], T. Piran[12], J. Rana[17], E. M. Sadler[4,10], L.P. Singer[18]

correspondence to: gh@astro.caltech.edu

**This PDF file includes:**

Materials and Methods
Supplementary Text
Figures S1-S5
Table S1
References (36-65)

## 1. Radio Data Reduction

### *1.1. VLA*

Radio observations of the EM170817 field were carried out with the Karl G. Jansky Very Large Array (*36*) in its C, CnB, and B configurations, under our target of opportunity programs (VLA/16A-206; PI: A. Corsi; VLA/17A-374; PI: K. Mooley). All observations reported here were carried out in *C*-band (nominal center frequency of 6 GHz, with a bandwidth of 4 GHz and *S*-band (nominal center frequency of 3 GHz, with a bandwidth of 2 GHz) with the Wideband Interferometric Digital Architecture (WIDAR) correlator. We used QSO J1258-2219 (*C*-band) and QSO B1245-197 (*S*-band) as our phase calibrator sources, and 3C 286 or 3C 147 as flux and bandpass calibrators.

VLA data were calibrated and flagged for radio frequency interference (RFI) using the VLA automated calibration pipeline which runs in the Common Astronomy Software Applications package (CASA, *37*). When necessary, additional flags were applied manually after calibration. Images of the observed field were formed using the CLEAN algorithm (*38*), which we ran in the interactive mode. When deconvolving bright background sources, care was taken not to include any clean components near the emission of EM170817.

Observations carried out after 2017 September 1 UTC were intermittently affected by rapidly fluctuating phases (timescale ~ minutes) caused by the on-going enhanced solar activity and potentially related activity in the troposphere. To mitigate phase errors in the *S*-band, a single round of phase-only self-calibration (including all sources brighter than 50 μJy in the initial model) was performed. At *C*-band, uncorrected phase errors resulted in a lower amplitude for all sources in the field of EM170817 during the first week of September. We therefore calculated a correction factor to the measured flux densities of the radio transient by estimating the average fractional change in the flux density of unresolved background radio sources in the field for each affected epoch. These corrections amount to ~30% of the flux for the most affected epoch (UT 2017 September 5) and to ~17% of the flux on the date of first discovery in C-band (UT 2017 September 3). A revised strategy of rapid phase calibration was implemented for observations in both bands on and after UT 2017 September 7, which greatly reduced the impact of the solar activity.

The results of our VLA follow-up campaign of EM170817 are reported in Table S1. Flux measurement errors are calculated as the quadratic sum of the map root-mean-square (rms) noise plus a 5% fractional error on the measured flux which accounts for inaccuracies in the flux density calibration. Where necessary, an amplitude correction factor is applied for known phase error issues in *C* band data. For non-detections, upper-limits are calculated as the measured flux at the position of EM170817 plus 2× the map rms. An example map with a high signal-to-noise detection (11σ) of EM170817 is shown in Figure S1.

### *1.2. ATCA*

We observed EM170817 on 2017 August 18, 21, 28 and September 05 using the Australia Telescope Compact Array (ATCA) under a target of opportunity program (CX391; PI: T. Murphy). During the Aug 18 observation the array was in the EW352 configuration, for all other observations it was in the 1.5A configuration. The Aug 18, 21 and 28 observations used two 2 GHz frequency bands with central frequencies of 8.5 and 10.5 GHz, while for the Sept 05 observation we centered these two frequency bands on 5.5 and 9.0 GHz. For all epochs, the flux scale and bandpass response was determined using the ATCA primary calibrator PKS B1934-638, and observations of QSO B1245-197 were used to calibrate the complex gains. The visibility data were reduced using the standard routines in the MIRIAD environment (*39*). For the August 18 epoch we noted a systematic error in the flux calibration of QSO B1245-197 and therefore scaled the flux densities for this epoch using the values listed for QSO B1245-197 in the ATCA Calibrator Database V3.

The calibrated visibility data from the August observations were then inverted and cleaned using the MIRIAD tasks INVERT, CLEAN and RESTOR. For these observations, we modelled and removed the host galaxy, NGC 4993, using the CASA tool IMFIT to fit a single Gaussian to this source, constraining the center of the Gaussian to be at the prior known location of NGC 4993 with a size identical to the restoring beam. In the resulting residual image, we then used the CASA task IMSTAT to place 3σ limits on the emission of EM170817 by measuring the rms within a region with a size six times the restoring beam and centered on its location.

For our observation in September, the calibrated visibility data were then split into the separate bands (5.5 GHz and 9.0 GHz), averaged to 32 MHz channels, and imported into DIFMAP (*40*). Bright field sources were modeled separately for each band with a combination of point-source and Gaussian components with power-law spectra. Care was taken not to include any components near the emission of EM170817 to avoid false detection. The residual images from each band were then averaged to form a wide-band image centered at 7.25 GHz. Restored images for each band were also generated by convolving the model components with the restoring beam and then averaged to form a wide-band image. Using a fit region, with size equal to the resolving beam and centered on the known location of EM170817, we measure a 4.1σ radio peak of 25 +/-6 μJy in this wide-band image.

### *1.3. GMRT*

We carried out observations of the EM170817 field with the Giant Metrewave Radio Telescope (GMRT) at 400 MHz, 700 MHz and 1.2 GHz under the Director's Discretionary Time (DDT) program (DDTB284; PI: K. De). Our first observation at 610 MHz was carried out with the GMRT Software Backend (GSB; *41*), with 32 MHz bandwidth. These data were calibrated, RFI flagged and imaged using the SPAM pipeline (*42*). All other observations, centered at 400 MHz (200 MHz bandwidth), 700 MHz and 1200 MHz (400 MHz bandwidth for both) were carried out using the upgraded GMRT Wideband Backend (GWB; *43*) to obtain the highest available sensitivities. For all observations, the pointing was centered at the location of the optical transient, 3C 283 was used as the complex gain calibrator and 3C 286 as the absolute flux scale and bandpass calibrators. These data were calibrated and RFI flagged using a custom-developed CASA pipeline. The data were then imaged interactively with the CASA task CLEAN, while incorporating a few iterations of phase-only self-calibration.

### *1.4. VLITE*

The VLA Low Band Ionosphere and Transient Experiment (VLITE; *44*) is a parallel observing system on the VLA that records the 340-384 MHz data from the prime focus P band receivers for nearly all observing programs. We report here VLITE results associated with the VLA target of opportunity programs described above. VLITE data are processed through a custom calibration and imaging pipeline which incorporates algorithms from the AIPS (*45*) and Obit (*46*) data reduction packages. The pipeline runs daily and initially sorts all data by the primary VLA Cassegrain primary observing band. For each data set, it then uses standard tasks to perform two cycles of automated excision of RFI, delay calibration, bandpass correction, and flux calibration.

All primary calibration uses a set of seven frequently observed standard calibrators, which may be observed at that primary observing band at any time during the day. Following the calibration, the pipeline performs wide-field, wide-bandwidth imaging of each target with up to two rounds of phase-only self-calibration and one round of amplitude and phase self-calibration.

The data were processed through the standard VLITE pipeline. The self-calibrated data were then re-imaged by hand using the Obit task 'MFImage'. An RFI-free band of 34 MHz centered at 338.7 MHz was chosen, and only baselines longer than 0.5 kλ were included, where λ is the observation wavelength. An additional phase-only self-calibration followed by an amplitude and phase self-calibration were performed. When needed, the nearby bright source (3C 283) was calibrated separately in amplitude and phase, and subtracted from the data (peeled) to reduce artifacts from the source. On 25 Aug. some of the observations in our 17A-374 program did not directly target the source but fall within the large VLITE field of view in eight separate images from that program. We convolved those eight images to a common beam and combined them using Obit mosaic utility tasks and a primary beam model derived from VLITE archival data. For days when multiple observations of the target were made at different primary observing bands, the visibility data were combined and imaged jointly.

### *1.5. GBT*

NGC 4933 was observed in the 21-cm line of neutral hydrogen on 2017 September 11 using the 100 m Robert C. Byrd Green Bank Telescope (GBT), under proposal GBT17B_395 (PI: F. Lockman) at an angular resolution of 9.1 arcminutes. Spectra were taken using the standard GBT L-band receiver (*47*) while the telescope was switched between the position of the galaxy and a nearby reference position for about 1 hour. The system was checked using observations of a known galaxy. A third order polynomial was fit to the final spectrum to remove a residual instrumental baseline. No 21-cm emission was detected at or near the expected velocity of 2900 km $s^{-1}$ *(48)* (see Figure S2) to an rms noise level of 7.0 mK (3.5 mJy) in a channel of 1.51 km $s^{-1}$. We translate that into an approximate limit on the neutral hydrogen mass of NGC 4993 by assuming a velocity width of 200 km $s^{-1}$, finding a 5σ mass limit of $<1 \times 10^8$ $M_\odot$.

## 2. Afterglow Modeling

The radio light curves were calculated using a numerical code described in (*49*). The output light curves have been found to be largely consistent with light curves produced by the BOXFIT code (*50*). In short, our code approximates the jetted blast wave at any lab time as a single zone emitting region which is a part of a sphere with an opening angle, $\theta_i$. The hydrodynamics, including the shock location and velocity and the jet spreading, are described in detail in (*51*). The hydrodynamic variables in the emitting region are taken as those that are immediately behind the shock. The emission from each location along the shock is calculated using standard afterglow theory (*25*), where the microphysics is parameterized by the fraction of internal energy that goes to the electrons, $\varepsilon_e$, the fraction of internal energy that goes to the magnetic field, $\varepsilon_B$, and the power-law index of the electron distribution, $p$. In all the models we used $\varepsilon_e = 0.1$ and $\varepsilon_B = 0.01$. In models that are consistent with the radio observations we used $p = 2.1$, so the models are also roughly consistent with the X-ray observations (*11,30*). In models that are inconsistent with the available radio observations we used the more typical value of $p = 2.4$. The code calculates the rest frame emissivity at any time and any location along the shock and the specific flux observed at any given viewing angle at any given time and frequency is then found by integrating the contribution over equal-arrival-time surfaces, with a proper boost to the observer frame and taking into account the light travel time.

## Supplementary Text

## 3. Independent Constraints on the Environment of GW170817

We can attempt to infer the local ISM environment of the merger from observational results on the global or local ISM for NGC 4993. However, the low star formation rate (SFR) of NGC 4993 and the lack of direct measures of the ISM at the position of EM170817 make this problematic (*5*). Considering our upper limit on the mass of neutral hydrogen, $M_{HI}$ of < 1 × $10^8$ $M_\odot$, as well as previous constraints (*52*), and assuming that the hydrogen is distributed uniformly over twice the stellar extent of the galaxy (estimated to be 18 × 16 kpc) implies a surface density of < 0.1 $M_\odot/pc^2$. Translating that into a local number density gives $n_{H\,I}$ < 0.04 $cm^{-3}$, with the total number density at most a factor of a few higher (assuming a face-on disk with scale-height of 100 pc; if the distribution is more spherical then it will lead to a lower limit). Repeating this exercise with a Sersic H I distribution gives a similar limit. We can invert this exercise, and look at the implied H I surface density from the estimated SFR. If the SFR is ~$10^{-2}$ $M_\odot$/yr (*5*) and assuming the same stellar extent as above, we can translate that into a star formation rate density and use the Kennicutt-Schmidt law (*53*, with n=1.4) to estimate $M_{H\,I}$ ~$10^7$ $M_\odot$, a factor of 10 lower than above. This would imply $n_{H\,I}$ ~ 0.004 $cm^{-3}$.

## 4. Comparison of 6 GHz Data with Models

The data obtained at 6 GHz, particularly before September 7, is subject to larger errors associated with rapidly fluctuating phases than data collected at 3 GHz. This is primarily because there is insufficient flux present in the EM170817 field at 6 GHz for robust self-calibration. However, it is still instructive to compare the observed light curve with the models (Figure S3). With the exception of one outlier point, the data are consistent with either a cocoon or an off-axis jet. The measured flux density on September 03 lies substantially above the predicted flux from the models. This observation was taken at a very low elevation (15 – 25°) resulting in sidelobe flux from the host galaxy AGN contaminating the location of the radio source. This may account for the higher flux density in this epoch, although intrinsic variability or scintillation cannot be ruled out.

Combining data at 3 and 6 GHz, we can also place constraints on the spectral slope of the radio emission. Using data taken on September 7 at 6 GHz and September 8 at 3 GHz, i.e., 1 day apart, we construct a 2-point spectral energy distribution (SED) for the radio counterpart to EM170817. Figure S3 compares this 2-point SED with the predicted SED for each of the models at 22 days post-merger. The radio data are consistent with the model predictions of an optically thin spectrum between 3 and 6 GHz.

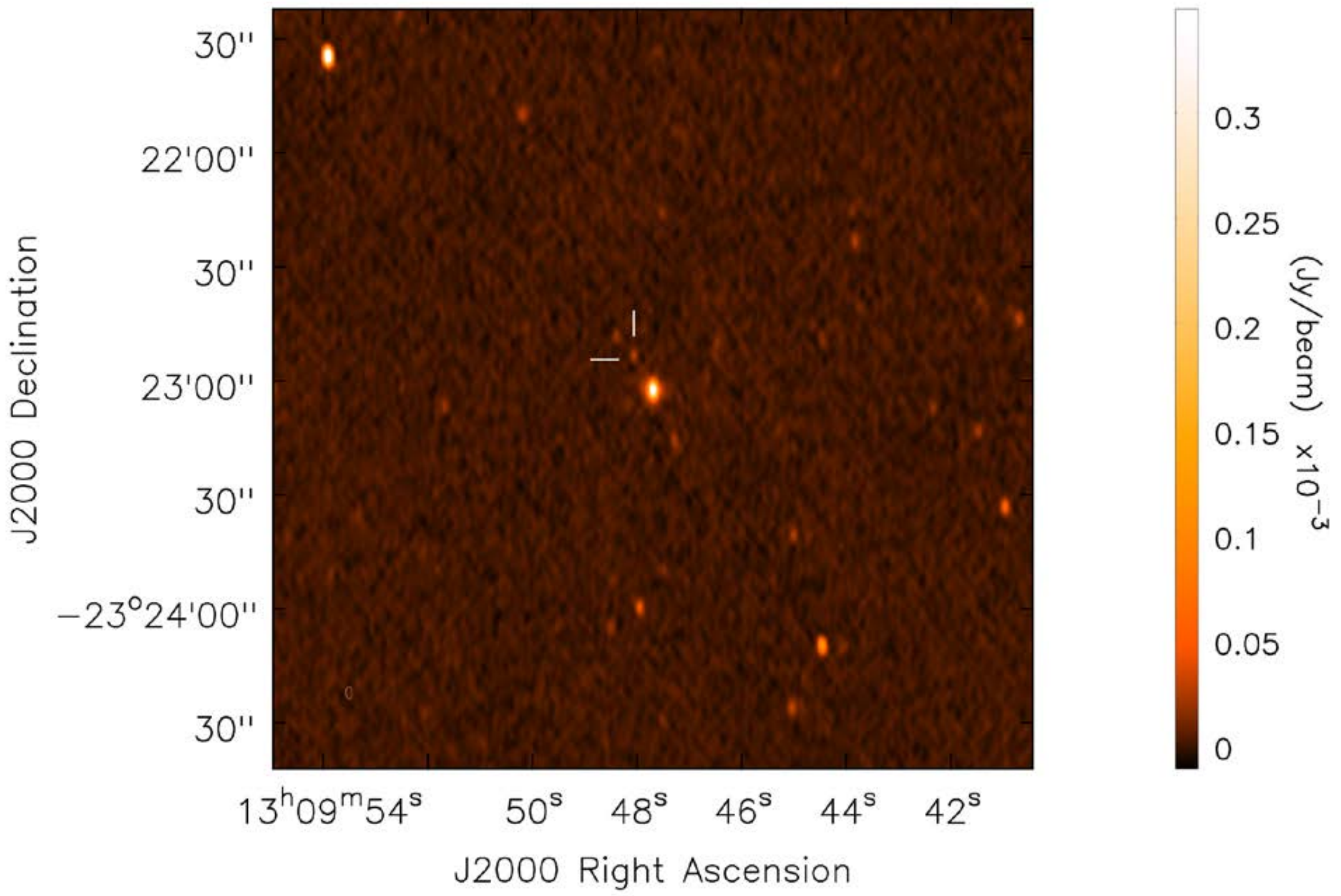

**Fig. S1. Deep radio image of EM170817.** A deep image of the ~3x3 arcminute field surrounding the location of NGC 4993 using VLA data at 3 GHz from 2017, September 8 and September 10. The AGN hosted by the galaxy is at the center of the image. The radio counterpart to EM170817 is highlighted and is detected at 25 +/- 2.2 μJy (11σ).

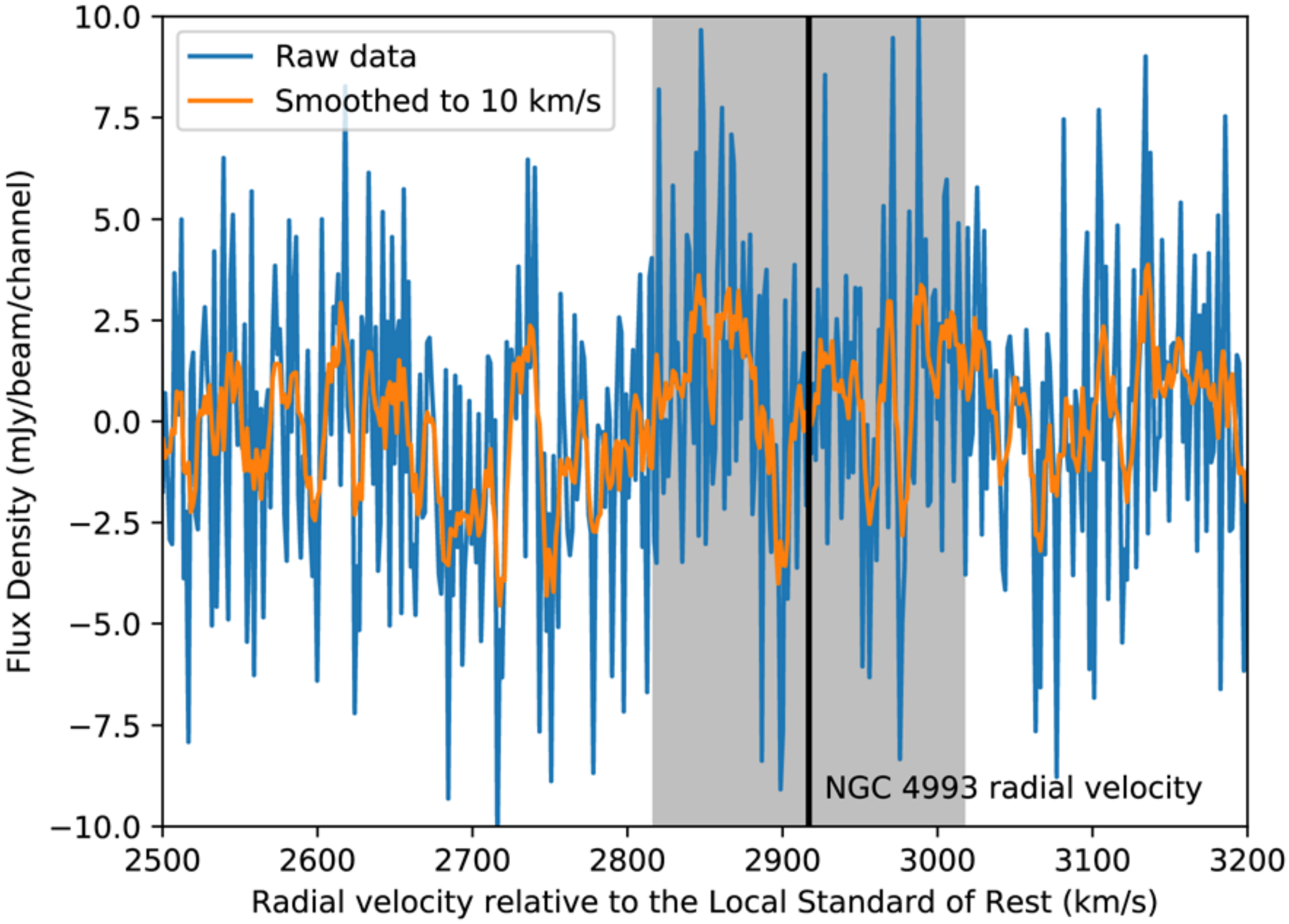

**Fig. S2. Green Bank Telescope spectrum of neutral hydrogen in the 21 cm line toward NGC 4993.** We plot the raw data in 1.5 km $s^{-1}$ channels, as well as a version smoothed to 10 km $s^{-1}$ channels, against radial velocity relative to the Local Standard of Rest. The radial velocity of NGC 4993 (*33*) is indicated with the vertical line, and the grey region spans a +/- 100 km $s^{-1}$ velocity width that we used to estimate an upper limit on the neutral hydrogen mass.

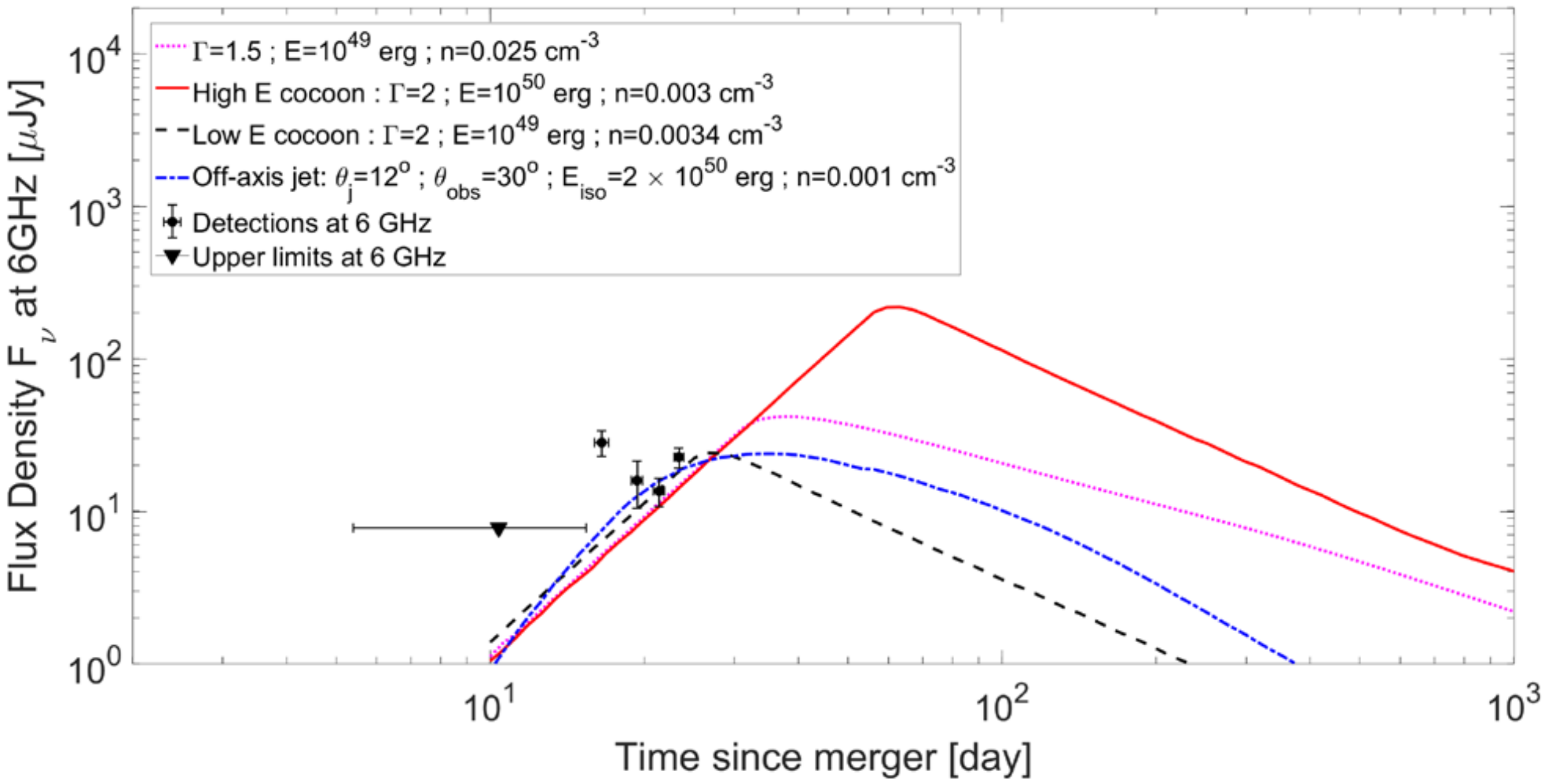

**Fig. S3. Radio Light Curve at 6 GHz.** We compare the light curve at 6 GHz with predictions from the four models previously shown in Fig. 3. With the exception of one outlier point, the data are consistent with either a cocoon or an off-axis jet. The flux density measurement on September 3 (16.5 days post-merger) was taken at very low elevation and is affected by sidelobe contamination from the host galaxy AGN.

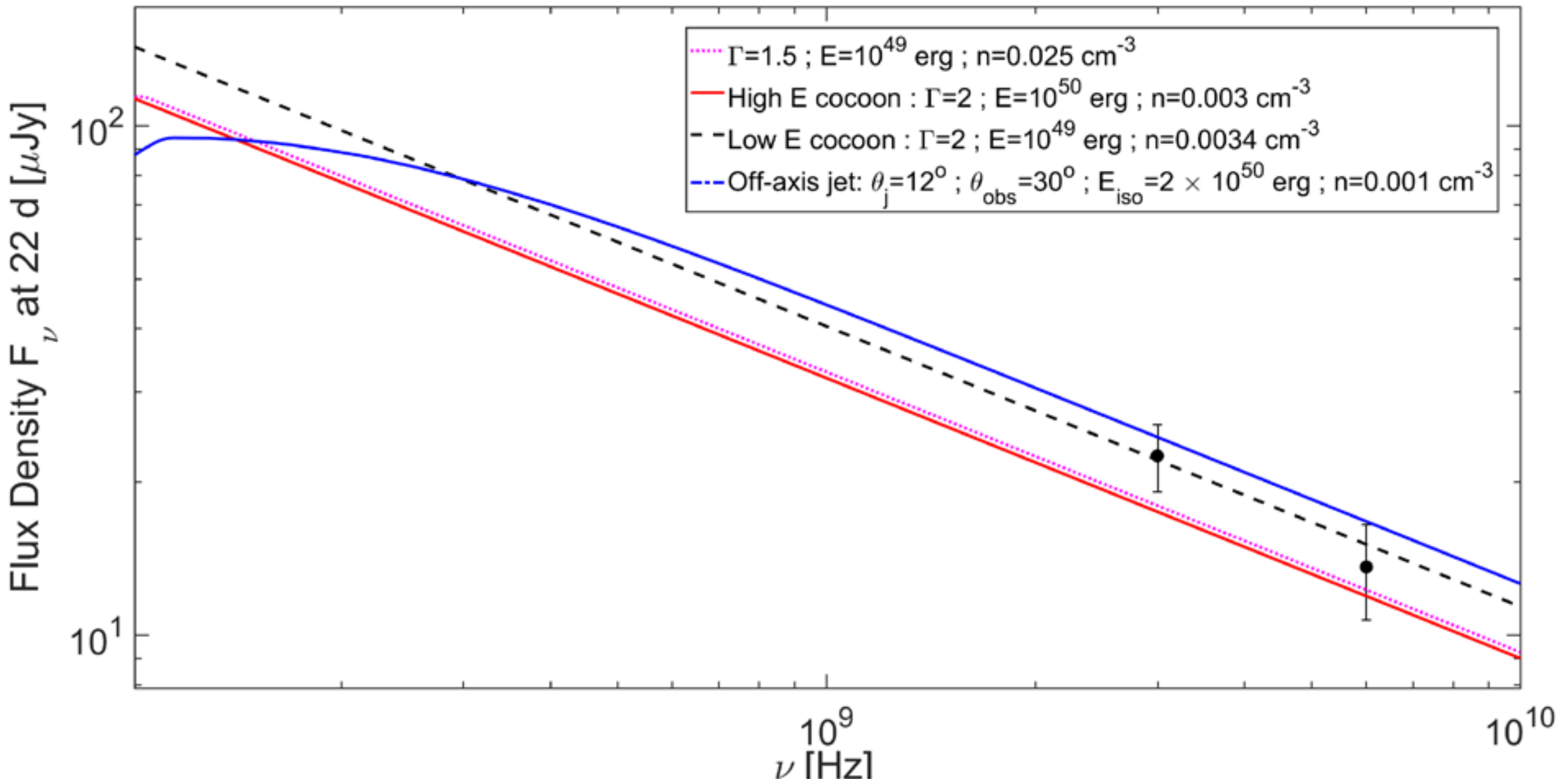

**Fig. S4. Radio Spectral Energy Distribution of EM170817.** We compare the data taken on September 7 at 6 GHz and September 8 at 3 GHz with the predicted radio spectrum for each of the models shown in Figure 3 at 22 days after the merger (September 8). The predicted spectrum is optically thin between these two frequency bands, which is consistent with the radio data.

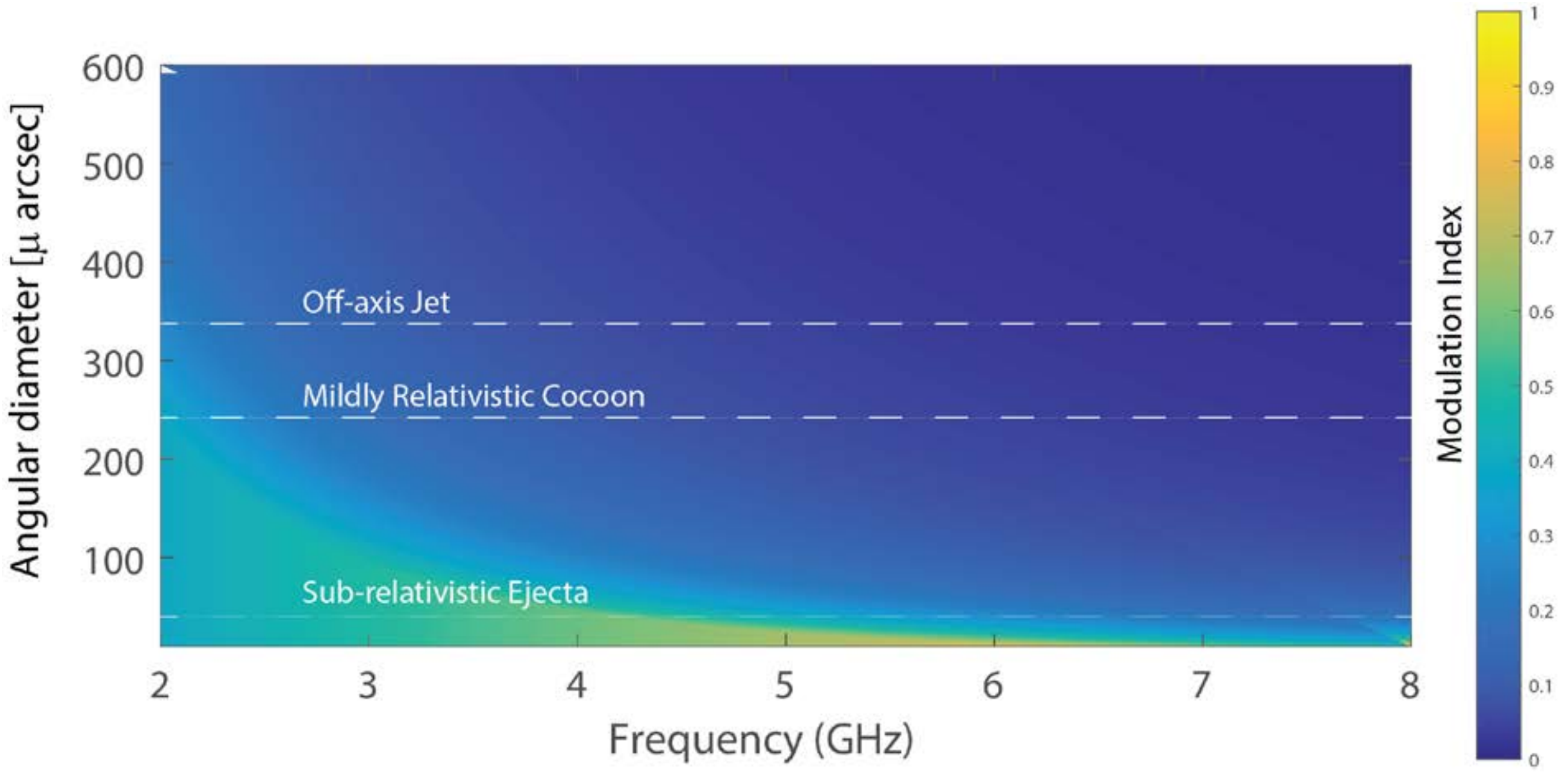

**Fig. S5. Predicted refractive scintillation of the radio source for different ejecta models.** The expected modulation index (fractional variation) is shown as a function of frequency and source size for refractive scintillation in the direction of EW170817. Horizontal dashed lines show the predicted source size at the time of the first detection (2017, September 2) for the various types of possible ejecta. Scintillation is expected to have had little impact on our observations, except in the unlikely case of radio emission produced by sub-relativistic ejecta, for which ~50% modulations at 3 GHz and 6 GHz would be present on 1 day timescales. This precludes using the light curve variability to establish source size, but does mean that the observed rise in the light curve (Figure 2, 3) is not contaminated by refractive scintillation.

**Table S1.** Compilation of all data collected as part of this campaign. All upper limits are 3 × RMS except for our VLA observations (excluding VLITE), which are reported as flux at optical location + 2 × RMS.

| UT Date | ΔT (d) | Telescope | ν (GHz) | Bandwidth (GHz) | $S_\nu$ (mJy) | Reference |
|---|---|---|---|---|---|---|
| Aug 18.21 | 0.68 | ATCA | 8.5 | 2.049 | < 0.120 | (*54*) |
| Aug 18.21 | 0.68 | ATCA | 10.5 | 2.049 | < 0.150 | (*54*) |
| Aug 18.46 | 0.93 | GMRT | 0.61 | 0.032 | < 0.195 | (*55)* |
| Aug 18.92 | 1.39 | VLA | 10 | 3.8 | <0.0154† | (*56*) |
| Aug 18.97 | 1.44 | VLITE/VLA | 0.3387 | 0.034 | < 34.8 | |
| Aug 19.95 | 2.42 | VLA | 6.2 | 4 | < 0.020 | (*57*) |
| Aug 19.95 | 2.42 | VLA | 9.7 | 4 | < 0.017 | (*58*) |
| Aug 19.95 | 2.42 | VLA | 15 | 6 | < 0.022 | (*59*) |
| Aug 19.97 | 2.44 | VLITE/VLA | 0.3387 | 0.034 | < 28.8 | |
| Aug 20.31 | 2.78 | GMRT | 0.4 | 0.2 | < 0.780 | (*60*) |
| Aug 20.46 | 2.93 | GMRT | 1.2 | 0.4 | < 0.098 | (*60*) |
| Aug 20.87 | 3.34 | VLA | 3 | 1.6 | < 0.032 | (*61*) |
| Aug 20.87 | 3.34 | VLITE/VLA | 0.3387 | 0.034 | < 44.7 | |
| Aug 21.23 | 3.7 | ATCA | 8.5 | 2.049 | < 0.135 | (*62*) |
| Aug 21.23 | 3.7 | ATCA | 10.5 | 2.049 | < 0.099 | (*62*) |
| Aug 22.88 | 5.35 | VLA | 6.2 | 4 | < 0.019 | (*63*) |
| Aug 25.90 | 8.37 | VLITE/VLA | 0.3387 | 0.034 | < 37.5 | |
| Aug 27.90 | 10.37 | VLA | 6.2 | 4 | 0.0078 ± 0.0026* | |
| Aug 28.18 | 10.65 | ATCA | 8.5 | 2.049 | < 0.054 | (*64*) |
| Aug 28.18 | 10.65 | ATCA | 10.5 | 2.049 | < 0.039 | (*64*) |
| Aug 29.45 | 11.92 | GMRT | 0.7 | 0.4 | < 0.123 | |
| Aug 30.98 | 13.45 | VLA | 6.2 | 4 | < 0.023 | |
| Aug 30.98 | 13.45 | VLITE/VLA | 0.3387 | 0.034 | < 20.4 | |
| Aug 31.46 | 13.93 | GMRT | 0.4 | 0.2 | < 0.600 | |
| Sep 01.89 | 15.37 | VLA | 6.2 | 4 | < 0.013 | |
| Sep 01.90 | 15.37 | VLITE/VLA | 0.3387 | 0.034 | < 11.4 | |
| Sep 02.89 | 16.36 | VLITE/VLA | 0.3387 | 0.034 | < 11.7 | |
| Sep 02.95 | 16.42 | VLA | 3 | 2 | 0.0187 ± 0.0063 | (*13*) |
| Sep 03.01 | 16.48 | VLA | 6.2 | 4 | 0.0283 ± 0.0054 | (*14)* |
| Sep 03.92 | 17.39 | VLA | 3 | 2 | 0.0151 ± 0.0039 | (*13*) |
| Sep 03.93 | 17.4 | VLITE/VLA | 0.3387 | 0.034 | < 6.9 | |
| Sep 04.86 | 18.33 | VLA | 3 | 2 | 0.0145 ± 0.0037 | |
| Sep 05.18 | 18.66 | ATCA | 7.25 | 4.098‡ | 0.025 ± 0.006 | (*15)* |
| Sep 05.52 | 18.99 | GMRT | 0.7 | 0.4 | < 0.140 | |
| Sep 05.88 | 19.35 | VLA | 6.2 | 4 | 0.0159 ± 0.0055 | |
| Sep 07.89 | 21.36 | VLA | 6.2 | 4 | 0.0136 ± 0.0029 | |
| Sep 07.89 | 21.36 | VLITE/VLA | 0.3387 | 0.034 | < 8.1 | |
| Sep 08.89 | 22.36 | VLA | 3 | 2 | 0.0225 ± 0.0034 | |
| Sep 08.90 | 22.37 | VLITE/VLA | 0.3387 | 0.034 | < 6.3 | |
| Sep 09.89 | 23.36 | VLITE/VLA | 0.3387 | 0.034 | < 4.8 | |
| Sep 09.89 | 23.36 | VLA | 6 | 4 | 0.0226 ± 0.0034 | |
| Sep 10.79 | 24.26 | VLA | 3 | 2 | 0.0256 ± 0.0029 | |
| Sep 10.88 | 24.35 | VLITE/VLA | 0.3387 | 0.034 | < 6.6 | |
| Sep 17.75 | 31.22 | VLA | 3 | 2 | 0.034 ± 0.0036 | (*65*) |

* Co-added from 22 Aug to Sep 1. Flux density consistent with marginal detection, but no discernible source present in the image.

† Value reflects our own analysis of these public data. A similar limit was reported in (*56*).

‡ 2×2049 MHz bands centred on 5.5 and 9 GHz